\documentclass[letterpaper, 10 pt, conference]{ieeeconf}  

\IEEEoverridecommandlockouts                              

\usepackage{cite}

\DeclareFontFamily{OT1}{pzc}{}
\DeclareFontShape{OT1}{pzc}{m}{it}{<-> s * [1.10] pzcmi7t}{}
\DeclareMathAlphabet{\mathpzc}{OT1}{pzc}{m}{it}

\let\proof\relax

\usepackage{amsmath,amssymb,amsfonts,amsthm}

\newtheoremstyle{assumptionstyle}
   {0pt}
   {0pt}
   {}
   {11pt}
   {\em}
   {\em}
   {5pt}
   {}
\theoremstyle{assumptionstyle}

\usepackage{algorithm}
\usepackage{algorithmic}

\newtheorem{lemma}{Lemma}

\usepackage{soul} 

\usepackage{xcolor}

\usepackage{cleveref} 

\usepackage{subcaption}
\usepackage{graphicx}
\usepackage{textcomp}
\usepackage{xcolor}
\def\BibTeX{{\rm B\kern-.05em{\sc i\kern-.025em b}\kern-.08em
    T\kern-.1667em\lower.7ex\hbox{E}\kern-.125emX}}

\usepackage{comment}

\title{\LARGE \bf
Gaussian Process-based Model Predictive Controller for Connected Vehicles with Uncertain Wireless Channel*
}

\author{Hassan Jafarzadeh $^{1}$ and Cody Fleming$^{2}$
\thanks{*This work was partially supported by the NSF under grants CPS-1739333.}
\thanks{$^{1}$Hassan Jafarzadeh is with the Department of Systems Engineering, University of Virginia, LinkLab-Olsson Hall, 151 Engineer’s Way, Charlottesville, VA 22903, USA {\tt\small hj2bh@virginia.edu}}%
\thanks{$^{2}$Cody Fleming is on the faculty of Mechanical Engineering at Iowa State University, 2078 Black Engr, 2529 Union Dr Ames, IA 50011-2030, USA {\tt\small flemingc@iastate.edu}}%
}

\begin{document}

\maketitle
\thispagestyle{empty}
\pagestyle{empty}

\begin{abstract}
In this paper, we present a data-driven Model Predictive Controller that leverages a Gaussian Process to generate optimal motion policies for connected autonomous vehicles in regions with uncertainty in the wireless channel. The communication channel between the vehicles of a platoon can be easily influenced by numerous factors, e.g. the surrounding environment, and the relative states of the connected vehicles, etc. In addition, the trajectories of the vehicles depend significantly on the motion policies of the preceding vehicle shared via the wireless channel and any delay can impact the safety and optimality of its performance. In the presented algorithm, Gaussian Process learns the wireless channel model and is involved in the Model Predictive Controller to generate a control sequence that not only minimizes the conventional motion costs, but also minimizes the estimated delay of the wireless channel in the future. This results in a farsighted controller that maximizes the amount of transferred information beyond the controller's time horizon, which in turn guarantees the safety and optimality of the generated trajectories in the future. To decrease computational cost, the algorithm finds the reachable set from the current state and focuses on that region to minimize the size of the kernel matrix and related calculations. In addition, we present an efficient recursive approach to decrease the time complexity of developing the data-driven model and involving it in Model Predictive Control. We demonstrate the capability of the presented algorithm in a simulated scenario.
\end{abstract}

\section{INTRODUCTION}
\label{sec:intro}
Connected, autonomous vehicles (CAVs) have the potential to identify, collect, process, exchange, and transmit real-time data leading to greater awareness of -- and response to -- events, threats, and imminent hazards within the vehicle's environment. At the core of this concept is a networked environment supporting very high speed interactions among vehicles (V2V), and between vehicles and infrastructure components (V2I)  to enable numerous safety and mobility applications.
When combined with automated vehicle-safety applications, vehicle connectivity provides the ability to respond and react in ways that drivers cannot, or cannot in a timely fashion, significantly increasing the effectiveness of crash prevention and mitigation. 
One application of CAVs involves Cooperative Adaptive Cruise Control (CACC), in which the following vehicle not only localizes the preceding vehicle via on-board sensing, but also receives information of the preceding vehicles' current state(s) through a wireless channel \cite{varaiya1993smart}. CACC is an extension to Adaptive Cruise Control that is capable of applying acceleration/deceleration depending on the existence of a preceding vehicle in a certain range. ACC is one of the first concepts enabling automation to involve a safety concern (i.e. time-to-collision) in its policy. 

Real-time control algorithms, e.g. Model Predictive Control (MPC), provide a foundation to significantly improve CACC performance. 
MPC involves generating a sequence of motion policies for some number of time steps in the future and executing a subset of this sequence in a receding fashion. Predicting the optimal trajectory into the future benefits the system in different ways, such as providing safety guarantees (e.g. obstacle avoidance) and minimizing control effort. Furthermore, in the context of CAVs, the predicted motion policy can be shared via a communication channel with the other vehicles. Being aware of future vehicle states allows better agility, maneuverability, safety, and optimality \cite{jafarzadeh2019learning}.

There is a growing literature on MPC in connected vehicle applications \cite{schmied2015nonlinear,stanger2013model}. However, most of this work assumes a perfect communication channel, meaning that an ego (following) vehicle receives packets from the preceding vehicle with no dropouts or delay. However, communication delays present a challenge, especially in vehicle-to-vehicle, or V2V, communication systems, which are characterized by a dynamic environment, high mobility, and low antenna heights/power on the transceivers (e.g. between vehicles and roadside units, telecommunications, etc)
\cite{viriyasitavat2015vehicular}. Communication delay can impact the performance of vehicle platoons in myriad ways. 
String stability is seriously compromised by the communication delay introduced by the network~\cite{liu2001effects}. 
Liu et al assume a constant communication delay (i.e. a very benign delay) and propose a simple fix to preserve string stability that is not optimal. The sub-optimality of their approach comes from the fact that controller designs do not explicitly account for delay in the communication links. In practice a wireless channel inevitably contains random delays, and packet losses vary based on the surrounding environment. Therefore, an approach of multiple time-varying delays was considered in \cite{petrillo2018adaptive}, which implements adaptive feedback gains to compensate for the errors originated from outdated packets, provides upper bound estimates for time delays, and  examines stability conditions using Linear Matrix Inequality techniques.

However, little work has been done on the effect of uncertainty of communication channel on the control policy of CAVs, and to our knowledge none have considered the wireless network itself as a dynamic variable. Typically, when the policy of a connected vehicle is optimized over a time horizon based on a given behaviour of the channel, it is assumed that the network will not vary by changing the state of the vehicle.  However, the states of the communicating vehicles are important factors that influence the wireless channel. 

States of the vehicle and wireless communication channel (e.g. PDR, or Packet Delivery Rate) are two interdependent variables, where changing either will affect the other \cite{jafarzadeh2019learning}. The ego vehicle should not only take into account the cost of control effort and its longitudinal distance from the preceding vehicle, but also the state of the wireless network should be involved in its controller model. For instance, the state of the ego vehicle where the communication channel has zero PDR is not desirable because at this state the ego vehicle will not receive any packet from the lead vehicle which in turn results in a sequence of conservative controls (because of the safety concerns) that impacts the overall performance. Accordingly, having a more stable wireless channel with a higher PDR improves the overall performance of the CAVs; however the challenge is, being data-driven, the wireless communication channel cannot be formulated directly as a straightforward mathematical model to be merged into the model predictive controller. 

To consider the PDR in MPC, we use Gaussian Processes (GP) which is a model-based method. GP is preferable in this work because the variations of the environment can be learned and used as a proxy model representing the behaviour of the wireless network for a given time horizon. Also, being a probabilistic method, the uncertainty can be incorporated into a long-term prediction (time horizon) by using GPs and the impact of the errors can be reduced \cite{schneider1997exploiting, kocijan2016modelling}. This is a data-efficient approach compared to the model-free methods (e.g. Q-learning or TD-learning) and can be used without additional interaction with the environment \cite{kamthe2017data}. After obtaining the proxy model, it can be used in MPC to generate a motion policy in which the PDR, control effort and the highway capacity usage are balanced in an optimal way. 
To treat the problem without building a proxy model, DMPC (Data-and  Model-Driven  Predictive  Control) can be applied, which is an iterative learning control approach. DMPC requires a feasible trajectory from an initial to terminal state, and iterates on the most recently generated full trajectory to find a better one until converging to the optimal trajectory~\cite{jafarzadeh2021dmpc}.

GPs have been used to model the dynamics of different systems, for instance leveraging Gaussian Processes to model cart-pole system and a unicycle robot \cite{deisenroth2011pilco} and inverted pendulum \cite{cutler2015efficient}. We use the same approach of \cite{kamthe2017data}, which is an improvement to Probabilistic Inference for Learning Control (PILCO) methodology initially presented in \cite{deisenroth2011pilco} that has the ability to propagate uncertainty through the time horizon of a predictive controller and learn the parameters of a LTI system. 

However, PILCO and its variants are computationally expensive, and in the context of CAVs it is not critical to learn a LTI system.
Another challenge with the majority of learning algorithms is that they often require too many trials to learn. For example, learning the mountain-car tasks often requires hundreds or thousands of trials, independent of whether using policy or value iterations, or policy search methods \cite{yahya2017collective, sutton2018reinforcement}. Thus the reinforcement learning algorithms have limited applicability to many real-world applications, e.g. mechanical systems, especially if their system dynamics change rapidly (such as wearing out quickly in low-cost robots) \cite{deisenroth2011pilco}. 

To decrease the size of the learned GP model in MPC, we use the concept of \textit{N-step reachable set} to focus on a small part of the model required to find the optimal motion policy from the current state. Also, we exploit the repetitiveness of the MPC to develop a recursive approach to decrease the computational burden of the GP in MPC. Furthermore, our approach learns the parameters of a wireless channel and integrates this information with known or explicit dynamical models.  Therefore, the contributions of this paper are 
\begin{enumerate}
    \item a computationally efficient controller that leverages and extends state-of-the-art learning algorithms, and 
    \item a control architecture for CAVs that accounts for communication uncertainty in a locally optimal way.
\end{enumerate}



\section{Background: MPC scheme for CAVs}\label{sec:background}
Consider a platoon of autonomous vehicles that generate their own optimal motion policy and share it with other vehicles in the platoon through a wireless communication channel.

Given a leader-following pair of vehicles, let  $x_t\in\mathbb{R}^{n}$ and $u_t\in\mathbb{R}^{m}$ be the state and control vectors of the ego, or following, vehicle at time step $t$. Its dynamical system is given by:  
\begin{equation}
    x_{t+1} = f\left(x_t,u_t \right). \label{eq:nldynamics}
\end{equation}
We assume that $f:\mathbb{R}^{n}\times\mathbb{R}^{m}\to\mathbb{R}^{n}$ is a smooth function. 
We first present the model predictive controller of the ego vehicle under the condition of perfect communication channel and then imperfect communication channel.

\subsection{Perfect Communication Channel}
In this case we assume that there is no delay in the wireless channel and the sent packet from the predecessor vehicle, that contains its predicted trajectory for $N$ time steps in the future, is being received by the ego vehicle at no time.
The controller of the ego vehicle for this case is proposed as an MPC architecture that solves the following finite-horizon optimization model with length of $N$ at time $t$: 
\begin{subequations}\label{eq:mpc-cavs-generic}
    \begin{align}
         \min_{{\textbf{u}}_t} & J = Q(x_{t+N|t}) + \sum_{k=t}^{t+N} h(x_{k|t},u_{k|t}) \label{eq:mpc_cost_function}\\
        \text{s.t.}\ & x_{k+1|t} = f\left(x_{k|t},u_{k|t} \right)  \label{eq:mpc-dynamic-const1}\\
        & x_{t|t} = x_t \label{eq:mpc-dynamic-const2} \\
        &\Phi(x_{k|t},x_{k|t}^p)\ge \varphi  \label{eq:mpc-dynamic-const3}\\
        & x_{k|t} \in \mathcal{X}, \quad u_{k|t} \in \mathcal{U}\quad \forall k \in \{t,\dots,t+N\}. \label{eq:mpc_dynamic_const4}
    \end{align}
\end{subequations}
where $h(x_{k|t},u_{k|t})$ is the stage cost and defined as:
\begin{equation}
    h(x_{k|t},u_{k|t})= \|{x}_{k|t}-{x}_{k|t}^p\|^2_{R_{1}} +\|{x}_{k|t}-{x}_k^{ref}\|^2_{R_{2}} + {\|u_{k|t}\|^2_{R_{3}}}
\end{equation}
and $x_{k|t}$ and $u_{k|t}$ are the state and control input vectors at step $k$ predicted at time $t$, respectively, which are shown as the solution of model (\ref{eq:mpc-cavs-generic}): 
\begin{equation}
  \label{eq:optimial_sol_MPC}
  \begin{gathered}
    \mathrm{\textbf{x}}_t=\left[x_{t+1|t}, x_{t+2|t},...,x_{t+N|t}\right]\\
    \mathrm{\textbf{u}}_t=[u_{t|t}, u_{t+1|t},...,u_{t+N-1|t}].
  \end{gathered}
\end{equation}

The receding horizon control law applies the first control input $u_{t|t}$ of $\mathrm{\textbf{u}}_t$ to shift the state of the system to $x_{t+1|t}$, and the process is repeated again from $t+1$. The cost function includes four terms to track: the state of predecessor vehicle ${x}_{k|t}^p$, a desired reference trajectory ${x}_k^{ref}$, minimal control effort, and minimum terminal cost $Q$. The tuning positive (semi)definite matrices are defined by $R_1$, $R_2$, and $R_3$. The optimal state vector of the lead vehicle, ${x}_{k|t}^p$, is sent via the wireless channel in the following packet
\begin{equation} \label{eq:leader_optimal_state_vector}
    \mathrm{\textbf{x}}_t^p=\left[x_{t+1|t}^p, x_{t+2|t}^p,...,x_{t+N|t}^p\right].
\end{equation}

The dynamical system and the initial condition are represented by (\ref{eq:mpc-dynamic-const1}) and (\ref{eq:mpc-dynamic-const2}). Constraint (\ref{eq:mpc-dynamic-const3}) enforces the safety conditions which can be defined simply as a minimum euclidean distance or as the \textit{Time To Collision} factor that cannot be less than a given parameter $\varphi$
\begin{equation} \label{eq:ttc}
  \Phi(x_{k|t},x_{k|t}^p) =
  \begin{cases}
  \infty                 & v_{k|t}^p\ge v_{k|t}\\
  \frac{d_{k|t}}{v_{k|t}-v_{k|t}^p} & v_{k|t}^p<v_{k|t}.
  \end{cases}
\end{equation}
$v_{k|t}$ and $d_{k|t}$ are the velocity of the ego vehicle and its euclidean distance from its predecessor at time step $k$ predicted at time $t$. It is assumed that there is a feasible trajectory from the initial state to the terminal state 
at time step $t$.

In this scheme (\ref{eq:mpc-cavs-generic}), it is assumed that the communication channel is perfect (common in literature, e.g. \cite{firoozi2018safe}), which means that the trajectory of the predecessor vehicle ${x}_{k|t}^p, \forall k \in \{t, \dots, N+t\}$ is transferred without any delay to the ego vehicle at the beginning of each time step, $t$. In this paper, this assumption will be relaxed to capture the uncertainty of the wireless communication channel.

\subsection{Imperfect Communication Channel}

The communication channel can be impacted by numerous factors such as relative states of the connected vehicles, surrounding infrastructure and vehicles, free space disturbances, etc. We define the predicted state vector of the communication channel established between the ego vehicle and its predecessor as
\begin{equation}
    \label{eq:packet_delivery_time_MPC}
    {\textbf{w}}_t=[{w}_{t+1|t}, {w}_{t+2|t}, \dots,{w}_{t+N|t}].
\end{equation}
$w_{t+k|t}$ shows the prediction for the packet delivery time of time step $k+t$ between two vehicles calculated at time $t$ (i.e. the inverse of packet delivery rate, PDR). The question that we address in this section is ``How does predicted delivery time affect the policy of the ego vehicle, and how can this delivery time be involved in the optimization model of the controller?" If the estimated delivery time $w_{t+k|t}$ is greater than the length of the time step used for the discretized dynamics in the general MPC formulation, there will be an estimated delay at time step $t+k$. In this case, the ego vehicle should use the most recent available packet, at time $t+k-1$, sent from the lead vehicle.

Due to the uncertainty in the delivery time of the packets, the ego vehicle can receive several packets at a single time step, $\{\mathrm{\textbf{x}}_{t_1}^p,\mathrm{\textbf{x}}_{t_2}^p,,...,\mathrm{\textbf{x}}_{t_m}^p\}$, where $t_1 < t_2 < ... < t_m \leqslant t+k $. Because the packet containing $\mathrm{\textbf{x}}_{t_m}^p$ is the most up-to-date (though not necessarily {\em current} at time $t+k$), it is used to calculate the motion policy of the ego vehicle. 
\begin{equation}  \label{eq:most_recent_packet}
    \mathrm{\textbf{x}}_{t_m}^p=[x_{t_m|t_m}^p,x_{t_m+1|t_m}^p, \dots, x_{t_m+N|t_m}^p].
\end{equation}
Nevertheless, at time step $t+k$, the first $t+k-t_m$ states in this packet belong to the past, are not useful anymore and should be neglected. Therefore, the useful packet shrinks to:
\begin{equation}  \label{eq:pruned_packet}
  \mathrm{\textbf{x}}_{t_m}^p=[x_{t+k|t_m}^p,x_{t+1|t_m}^p, \dots, x_{t_m+N|t_m}^p].
\end{equation}

The length of this packet determines the applicable time horizon in the model predictive controller that the ego vehicle can consider to involve the trajectory of the lead vehicle:
\begin{equation}  \label{eq:changing time horizon}
  \mathrm{N}_{t+k}=t_m+N-(t+k).
\end{equation}
where $\mathrm{N}_{t+k}$ denotes the length of time horizon at time $t+k$, $\mathrm{N}_{t+k} \leqslant N $. Increases in the value of $w_{t+k|t}$ decreases $t_m$, which in turn decreases the length of the useful time horizon $\mathrm{N}_{t+k}$. Fewer useful states from the lead vehicle causes the ego vehicle to take more conservative policies to satisfy the safety constraints, and this approach results in more cost for the entire whole trajectory; i.e. the controller might find local optima due to lack of longer-term information.

In this work we add awareness of the cost to physical system performance due to communication delays, which updates the cost function of model (\ref{eq:mpc-cavs-generic}) to:
\begin{equation} \label{eq:updated_cost_function_delay}
    \min_{{\textbf{u}}_t}  J = Q(x_{t+N|t}) + \sum_{k=t}^{t+N} [h(x_{k|t},u_{k|t}) + y(w_{k|t})],
\end{equation}
in which $y$ is a positive definite function. Adding a notion of the packet delivery time of the wireless channel to the cost function of the model equips the system with a farsighted controller to guarantee the availability of the packet with a maximum possible length at each time step in the future, ultimately resulting in a safer and more optimal motion policy. Therefore, the objective function includes the cost of the delay in communication channel for the trajectories that result in a low quality state of communication channel, $\omega_{t+k|t}$. Thus, the optimality condition forces the model to generate trajectories with desirable quality of the communication while satisfying the constraints. In the next section we use Gaussian Processes to estimate the state of wireless channel at each time step, $w_{k|t}$, as a function of the states of two connected vehicles.

\section{Gaussian Process for CAVs}\label{sec:gp-cavs}
In the following, we will present the main components of Gaussian Processes for connected autonomous vehicles and show how to involve it in model predictive control. Later in the section, an efficient recursive approach will be developed to improve the time complexity of the presented algorithm.

\subsection{Probabilistic Model for Wireless Channel}
Assume that the delay of packet delivery from the lead vehicle to the ego vehicle at time $t$ can be described by the following unknown function
\begin{equation}
    \label{eq:pdr_unknown_func}
    \omega_{t} = \Omega(x_{t} , x_{t}^p, e) + \varepsilon
\end{equation}
where $e$ show the external environment and $\varepsilon$ is noise $\varepsilon \sim \mathcal{N}(0,\sigma_{\varepsilon}^2)$. In this paper, we consider a Gaussian Process setting where we seek deterministic control inputs $\textbf{u}_{t}$ that minimize the expected long-term cost
\begin{equation}
    \label{eq:updated_obj_func}
    \min_{\textbf{u}_t} \Big\{ J_{t \rightarrow t+N} + \sum_{k=t}^{t+N} y(\mathbb{E}_{\omega_{k|t}}[\omega_{k|t}]) \Big\}.
\end{equation}

$J_{t \rightarrow t+N}$ denotes the conventional cost function in MPC, i.e. equation (\ref{eq:mpc_cost_function}), and $y(\mathbb{E}_{\omega_{k|t}}[\omega_{k|t}])$ denotes the cost of expected delay in packet delivery at time step $k$ calculated at time $t$. To implement the GP we use $\textbf{x} = [x, x^p]^T,  \textbf{x} \in \mathbb{R}^{2n}$ as training inputs. $x$ and $x^p$ indicate the state of ego vehicle and the state of lead vehicle associated with $x$, and \(\omega \in \mathbb{R}^{\geq 0}\) as training target. A GP as a probabilistic, non-parametric model can be fully specified by a mean function \(m(\cdot)\) and a covariance function \(k(\cdot,\cdot)\) which is defined as squared exponential (Radial Basis Function, RBF):
\begin{equation}
\label{eq:kernel_definition}
    k(\textbf{x}, \textbf{x}') = \sigma^2_{\Omega} exp \Big( -\frac{1}{2} (\textbf{x}-\textbf{x}')^T  \textbf{L}^{-1}  (\textbf{x} -\textbf{x}')  \Big).
\end{equation}
where \( \sigma^2_{\Omega} \) is signal variance, and \( \textbf{L} = diag([\ell_1^2, \dots, \ell_{2n}^2 ]) \) with length-scales \( \ell_1, \dots, \ell_{2n} \). 

Assuming that there are $r$ training inputs and corresponding training targets, we collect $\textbf{X} = [\textbf{x}_1, \dots, \textbf{x}_r]^T$ and $ \pmb{\omega} = [\omega_1, \dots, \omega_r]^T $. Given the test input denoted $\textbf{x}_*$, the posterior predictive distribution of $\omega_*$ is Gaussian $p(\omega_*|\textbf{x}_*, \textbf{X},\pmb{\omega}) = \mathcal{N}\big(\omega_*|m(\textbf{x}_*), \sigma^2(\textbf{x}_*) \big)$ where 
\begin{equation}
\label{eq:mean_test_data}
    m(\textbf{x}_*) = k( \textbf{X}, \textbf{x}_*)^T(\textbf{K} + \sigma^2_{\varepsilon}\textbf{I})^{-1}\pmb{\omega},
\end{equation}
\begin{equation}
\label{eq:variance_test_data}
    \sigma^2(\textbf{x}_*) = k( \textbf{x}_*, \textbf{x}_*) - k( \textbf{X}, \textbf{x}_*)^T(\textbf{K} + \sigma^2_{\varepsilon}\textbf{I})^{-1}k( \textbf{X}, \textbf{x}_*).
\end{equation}
$\textbf{K}$ is the Gram matrix with entries of $\textbf{K}_{i,j} = k(\textbf{x}_i,\textbf{x}_j)$ \cite{rasmussen2003gaussian}.

It is assumed that $ p(\textbf{x}_{k|t}) = \mathcal{N}(\textbf{x}_{k|t}|\pmb{\mu}_t, \pmb{\Sigma}_t )$, where $\pmb{\mu}_{k|t}$ and $\pmb{\Sigma}_{k|t}$ are the mean and covariance of $\textbf{x}_{k|t}$. From equation (\ref{eq:mean_test_data}), the cost function (\ref{eq:updated_obj_func}) can be rewritten as
\begin{equation}
    \label{eq:cost_func_finalupdate}
    \min_{\textbf{u}_{t}} J_{t \rightarrow t+N} + \sum_{k=t}^{t+N} y\big(k( \textbf{X}, \textbf{x}_{k|t})^T(\textbf{K} + \sigma^2_{\varepsilon}\textbf{I})^{-1}\pmb{\omega}\big),
\end{equation}
where $\textbf{x}_{k|t}$ is the aggregated vector including the state vectors of the ego and lead vehicles. The first $n$ elements of this vector (i.e. test input) in $\textbf{x} = [x, x^p]^T$ belongs to the ego vehicle and will be decided by the model predictive controller such that the overall cost takes a minimum value. The constraints (\ref{eq:mpc-dynamic-const1})-(\ref{eq:mpc_dynamic_const4}) hold for this cost function. After finding the best control input vector $\textbf{u}_{t}$, the first control action, $u_{t|t}$, is applied and the state of the ego vehicle is updated to $x_{t+1|t}$.

\subsection{Efficient Recursive GP-MPC} \label{sec:efficient_GPMPC}

The GP-MPC algorithm needs to invert the Gram matrix, $\textbf{K}$, in cost function (\ref{eq:cost_func_finalupdate}), which has time complexity of $O(r^3)$, where $r$ is the number of training data. For large training sets (ten thousands or more) construction of GP regression becomes an intractable problem. We improve the time complexity of the algorithm in this section, leveraging the concept of \textit{Reachable Set} to decrease the size of matrix $\textbf{K}$.

\textit{\textbf{Definition}} 1 (\textbf{one-step reachable set} $\mathcal{B}$): For the system (\ref{eq:nldynamics}), the one-step reachable set from the set $\mathcal{B}$ is denoted as
\begin{multline}
    Reach(\mathcal{B}) = \Big\{x \in \mathbb{R}^{n}: \exists x(0) \in \mathcal{B}, \ \exists u(0) \in \mathcal{U} \\ s.t. \ x=f(x(0), u(0))\Big\}
\end{multline}
$Reach(\mathcal{B})$ is the set of states that can be reached in one time step from state $x(0)$. $N$-step reachable set are defined by iterating $Reach(.)$ computations \cite{borrelli2017predictive}.

\textit{\textbf{Definition}} 2 (\textbf{$N$-step reachable set} $\mathcal{R}_N(\mathcal{X}_0)$): For a given initial set $\mathcal{X}_0 \subseteq \mathcal{X}$, the $N$-step reachable set $\mathcal{R}_N(\mathcal{X}_0)$ of the system (\ref{eq:nldynamics}) subject to constraints (\ref{eq:mpc-dynamic-const3}) and (\ref{eq:mpc_dynamic_const4}) is defined as \cite{borrelli2017predictive}: 
\begin{multline}
    \mathcal{R}_{t+1}(\mathcal{X}_0) = Reach(\mathcal{R}_{t}(\mathcal{X}_0)), \ \mathcal{R}_0(\mathcal{X}_0)= \mathcal{X}_0, \\ t=\{0, \dots, N-1 \} 
\end{multline}
Using the \textit{N-step Reachable Set} concept, we define sub-matrix $\Bar{\textbf{K}}(t)$ that is extracted from matrix $\textbf{K}$
\begin{equation} \label{eq:new_small_kernel_matrix}
    \Bar{\textbf{K}}_{i,j}(t) = \{k(\textbf{x}_i, \textbf{x}_j) | \textbf{x}_i, \textbf{x}_j \in \mathcal{R}_N(\textbf{x}_t) 	\cap \textbf{X} \}.
\end{equation}

This will result in the following cost function
\begin{equation}
    \label{eq:cost_func_final_reachable_set}
    \min_{\textbf{u}_{t}} J_{t \rightarrow t+N} + \sum_{k=t}^{t+N} y\big(k(\Bar{\textbf{X}}_t, \textbf{x}_{k|t})^T(\Bar{\textbf{K}}(t) + \sigma^2_{\varepsilon}\textbf{I})^{-1}\Bar{\pmb{\omega}}_t\big),
\end{equation}
where $  \Bar{\textbf{X}}_t = \{ \textbf{x} | \textbf{x} \in  \mathcal{R}_N(\textbf{x}_t) \cap \textbf{X}\} $ and $\Bar{\pmb{\omega}}_t$ is the associated training output. Assume that the sampling has been executed randomly and for each time step, $\nu$ samples are available on average. The number of training data extracted from the overall training data, $\textbf{X}$, would be $\nu N$, where $\nu N \ll r$. Constructing the sub-matrix $\Bar{\textbf{K}}(t)$ is straightforward and can be implemented based on the state vector of lead vehicle, $\textbf{x}^p$, and the defined safety constraint (\ref{eq:mpc-dynamic-const3}). 

\begin{lemma} \label{lemma:GPMPC_time_complexity}
Given Matrix $\Bar{\textbf{K}}(t-1)$ denoting the sub-matrix extracted from matrix $\textbf{K}$ and containing \textit{N-step Reachable Set} at time step $t-1$, and its inverse $\Bar{\textbf{K}}^{-1}(t-1)$, then $\Bar{\textbf{K}}^{-1}(t)$ can be calculated in $O(\nu ^3 N^2)$ time.  
\end{lemma}

\proof
In the given matrix $\Bar{\textbf{K}}(t-1)$, $\nu$ training data representing the wireless channel at time step $t-1$ should be removed because the controller has implemented one step of control input, $u_{t|t}$. In addition, training data representing time step $t+N$ that is obtained from $\mathcal{R}_N(\textbf{x}_t)$ should be added to the matrix. The result of these two steps will be matrix $\Bar{\textbf{K}}(t)$. According to our assumptions, $\nu$ training data will be removed and added, which keep the size of matrix $\Bar{\textbf{K}}^{-1}(t)$ the same, $\nu N$.

We use the Sherman–Morrison formula \cite{juarez2016relationship} to find $\Bar{\textbf{K}}^{-1}_{\Bar{1}, \Bar{1}}(t-1)$ (i.e. matrix $\Bar{\textbf{K}}^{-1}(t-1)$ that its first column and row are removed) from $\Bar{\textbf{K}}^{-1}(t-1)$ remove the first column and the first row from matrix $\Bar{\textbf{K}}(t-1)$. Based on Sherman-Morrison formula 
\begin{equation} \label{eq:sherman_morrison_update}
    \Bar{\textbf{K}}^{-1}_{\Bar{1}, \Bar{1}}(t-1) =  \Big( \Bar{\textbf{K}}(t-1) - pq^T\Big)^{-1}_{\Bar{1}, \Bar{1}}. 
\end{equation}
where $p$ and $q$ are defined as: $p = \Bar{\textbf{K}}_1(t-1) - e_1$, and $q = e_1$, ($e_i$ is $i^{th}$ canonical column vector). The right hand side of equation (\ref{eq:sherman_morrison_update}) can be expanded as: 
\begin{multline} \label{eq:sherman_morrison_update_1}
    \Big( \Bar{\textbf{K}}(t-1) - pq^T\Big)^{-1}_{\Bar{1}, \Bar{1}} = \\ \Bar{\textbf{K}}^{-1}(t-1) + \frac{\Bar{\textbf{K}}^{-1}(t-1)  pq^T \Bar{\textbf{K}}^{-1}(t-1)}{1 - q^T \Bar{\textbf{K}}^{-1}(t-1)p}.
\end{multline} 
$1 - q^T \Bar{\textbf{K}}^{-1}(t-1)p$ is assumed to be invertible. This term can be calculated in $O(\nu^2N^2)$, which comes from multiplication of the square matrix $\Bar{\textbf{K}}^{-1}(t-1)$ with dimension $(\nu N \times \nu N)$ and vector $p$ and $q^T$. We denote the new training set as $ \Bar{\textbf{X}}'_{t-1}$, that has one less training data than $ \Bar{\textbf{X}}_{t-1}$.

In the second step, a new training data, $\textbf{x}$, is added. 
\begin{equation}
M:=
    \begin{bmatrix} \label{eq:matrix_updates}
        \Bar{\textbf{K}}_{\Bar{1}, \Bar{1}}(t-1) & k( \Bar{\textbf{X}}'_{t-1}, \textbf{x})\\
        k( \Bar{\textbf{X}}'_{t-1}, \textbf{x})^T & k( \textbf{x}, \textbf{x})
\end{bmatrix}
\end{equation}
Now, given $\Bar{\textbf{K}}^{-1}_{\Bar{1}, \Bar{1}}(t-1)$, we can find the inverse of matrix $M$ by using Schur complement \cite{zhang2006schur}. The Schur's complement of $\Bar{\textbf{K}}_{\Bar{1}, \Bar{1}}(t-1)$ in matrix $M$ is given by:
\begin{equation} \label{eq:schur_complement}
    M/\Bar{\textbf{K}}_{\Bar{1}, \Bar{1}}(t-1) := k( \textbf{x}, \textbf{x}) - k( \Bar{\textbf{X}}'_{t-1}, \textbf{x})^T \Bar{\textbf{K}}^{-1}_{\Bar{1}, \Bar{1}}(t-1) k( \Bar{\textbf{X}}'_{t-1}, \textbf{x})
\end{equation}
Assuming that $M/\Bar{\textbf{K}}_{\Bar{1}, \Bar{1}}(t-1)$ is invertable, $M^{-1}$ can be calculated in $O(\nu^2N^2)$ as:
\begin{equation} \label{eq:M_inverse_2_2}
    M^{-1}_{2,2} = (M/\Bar{\textbf{K}}_{\Bar{1}, \Bar{1}}(t-1))^{-1}
\end{equation}
\[M^{-1}_{2,1} = -M^{-1}_{2,2}  k( \Bar{\textbf{X}}'_{t-1}, \textbf{x})^T \Bar{\textbf{K}}^{-1}_{\Bar{1}, \Bar{1}}(t-1)\]
\[ M^{-1}_{1,2} = -\Bar{\textbf{K}}^{-1}_{\Bar{1}, \Bar{1}}(t-1)  k( \Bar{\textbf{X}}'_{t-1}, \textbf{x}) M^{-1}_{2,2}\]
\begin{equation} \label{eq:M_inverse_1_1}
    M^{-1}_{1,1} =  \Bar{\textbf{K}}^{-1}_{\Bar{1}, \Bar{1}}(t-1) - M^{-1}_{1,2} k( \Bar{\textbf{X}}'_{t-1}, \textbf{x})^T \Bar{\textbf{K}}^{-1}_{\Bar{1}, \Bar{1}}(t-1).
\end{equation}
Removing and adding a row and a column both are calculated in $O(\nu^2N^2)$. These two steps should be repeated for $\nu$ times, which yields to time complexity of $O(\nu^3N^2)$.

Details of the algorithm are presented in Algorithm (\ref{alg}).
\begin{algorithm}
\caption{: Calculating $\Bar{\textbf{K}}^{-1}(t)$ from $\Bar{\textbf{K}}^{-1}(t-1)$}
\label{alg}
\begin{algorithmic}[1]
\STATE{load training data $\textbf{X}$, $\Bar{\textbf{K}}(t-1)$ and $\Bar{\textbf{K}}^{-1}(t-1)$}
\STATE{ $\Bar{\textbf{X}}_t \leftarrow \mathcal{R}_N(\textbf{x}_t) 	\cap \textbf{X}$}
\STATE{$\Bar{\textbf{X}}'_{t-1} \leftarrow \Bar{\textbf{X}}_{t-1}$}
\STATE{calculate $\Bar{\textbf{K}}(t)$, equation (\ref{eq:new_small_kernel_matrix})}
\STATE{$E \leftarrow \Bar{\textbf{K}}(t-1)$, $E^{-1} \leftarrow \Bar{\textbf{K}}^{-1}(t-1)$}
\FOR{$i = 1:\nu$}
\STATE{$p \leftarrow E - e_1$, and $q = e_1$}
\STATE{$E^{-1}_{\Bar{1}, \Bar{1}}\leftarrow  (E - pq^T)^{-1}_{\Bar{1}, \Bar{1}}$, equations (\ref{eq:sherman_morrison_update}),( {\ref{eq:sherman_morrison_update_1}})}
\STATE{$\Bar{\textbf{X}}'_{t-1} \leftarrow \Bar{\textbf{X}}'_{t-1} - \Bar{\textbf{X}}'_{t-1}\{1\}$}
\STATE{load new training data $\textbf{x}$ from $\textbf{X}$}
\STATE{$E \leftarrow \begin{bmatrix} \label{eq:matrix_updates1}
        E_{\Bar{1}, \Bar{1}} & k( \Bar{\textbf{X}}'_{t-1}, \textbf{x})\\
        k( \Bar{\textbf{X}}'_{t-1}, \textbf{x})^T & k( \textbf{x}, \textbf{x})
\end{bmatrix}$}
\STATE{calculate $k( \Bar{\textbf{X}}'_{t-1}, \textbf{x})$ and $k( \textbf{x}, \textbf{x})$}
\STATE{$S \leftarrow k( \textbf{x}, \textbf{x}) - k( \Bar{\textbf{X}}'_{t-1}, \textbf{x})^T E^{-1}_{\Bar{1}, \Bar{1}} k( \Bar{\textbf{X}}'_{t-1}, \textbf{x})$}, equation (\ref{eq:schur_complement})
\STATE{$M^{-1}_{2,2} \leftarrow S^{-1}$}
\STATE{$M^{-1}_{2,1} \leftarrow -M^{-1}_{2,2}  k( \Bar{\textbf{X}}'_{t-1}, \textbf{x})^T E^{-1}_{\Bar{1}, \Bar{1}}$}
\STATE{$M^{-1}_{1,2} \leftarrow -E^{-1}_{\Bar{1}, \Bar{1}}  k( \Bar{\textbf{X}}'_{t-1}, \textbf{x}) M^{-1}_{2,2}$}
\STATE{$M^{-1}_{1,1} \leftarrow  E^{-1}_{\Bar{1}, \Bar{1}} - M^{-1}_{1,2} k( \Bar{\textbf{X}}'_{t-1}, \textbf{x})^T E^{-1}_{\Bar{1}, \Bar{1}}$}
\STATE{$E^{-1} \leftarrow \begin{bmatrix} \label{eq:matrix_updates2}
        M^{-1}_{1,1} & M^{-1}_{1,2}\\
        M^{-1}_{2,1} & M^{-1}_{2,2}
\end{bmatrix}$}
\STATE{$\Bar{\textbf{X}}'_{t-1} \leftarrow \Bar{\textbf{X}}'_{t-1} \cup \textbf{x}$}
\ENDFOR
\STATE{$\Bar{\textbf{K}}^{-1}(t) \leftarrow E^{-1}$}
\STATE{$\Bar{\textbf{K}}(t) \leftarrow E$}
\STATE{$\Bar{\textbf{X}}_t \leftarrow \Bar{\textbf{X}}'_{t-1}$}
\end{algorithmic}
\end{algorithm}


\section{Example Scenario - Uncertain Channel}
\label{sec:example}
\begin{figure}
    \centering
    \includegraphics[width=0.43\textwidth]{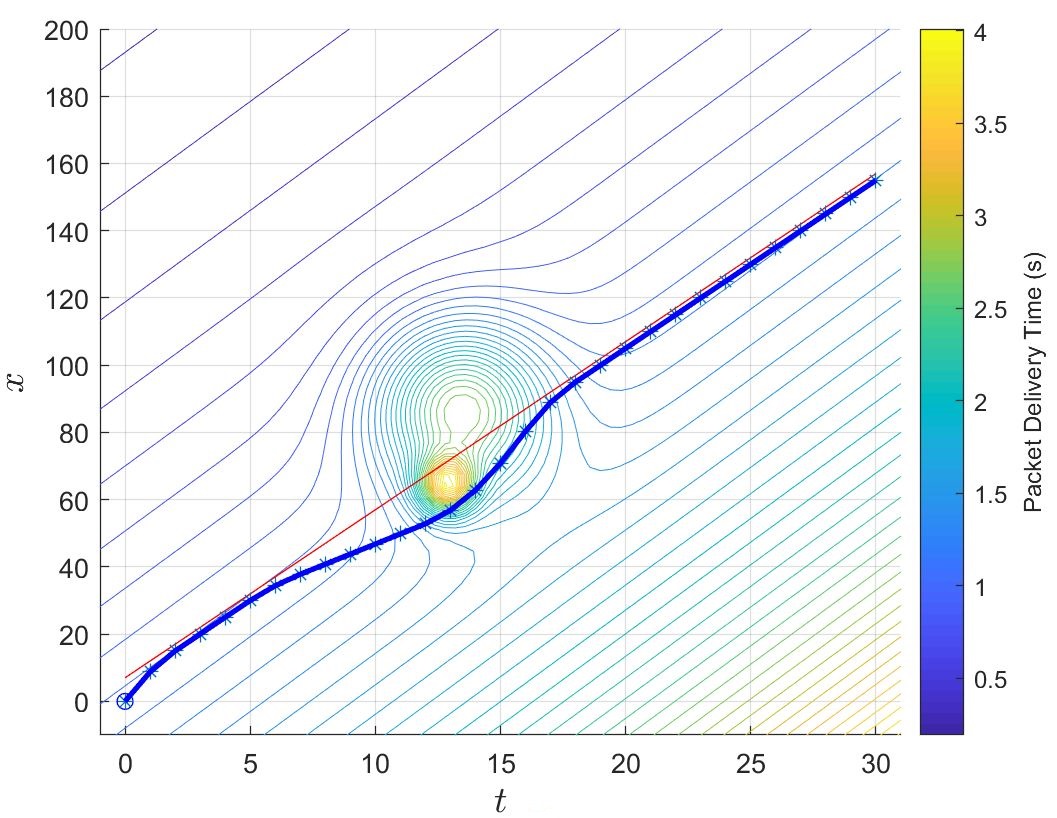}
    \caption{Optimal trajectory of ego vehicle generated by GP-MPC to avoid a region with a considerable delay time in packet delivery.}
    \label{fig:traj}
\end{figure}

In this section, we apply the method on two connected vehicles and assume that there is a region where the packet delivery rate is impacted significantly; this region is unknown a priori by the controller. The expected packet delivery time of the wireless channel between two connected vehicles is demonstrated by contour plot in Figure (\ref{fig:traj}). Wireless channel model is learned by Gaussian Processes and according to Algorithm (\ref{alg}) in section (\ref{sec:efficient_GPMPC}), and the reachable set of the current state of the ego vehicle is calculated in the usual way,  $\mathcal{R}_N(\textbf{x}_t)$. After obtaining the necessary training input for state $\textbf{x}_t$, $\Bar{\textbf{X}}_t$, the inverse of associated kernel matrix $\Bar{\textbf{K}}^{-1}(t)$ is calculated according to Lemma (\ref{lemma:GPMPC_time_complexity}). Finally, the cost function (\ref{eq:cost_func_final_reachable_set}) is built, and the GP-MPC proceeds by solving a sequential quadratic program.

It is assumed that the lead vehicle has a constant velocity $5 m/s$ and the ego vehicle should keep the safe distance of at least $\varphi = 2 m$ form the lead vehicle, $\Phi(x_{k|t},x_{k|t}^p)\ge 2$, which is enforced by constraint (\ref{eq:mpc-dynamic-const3}). The time horizon is considered to be $N=10$ and the length of each time step is $dt = 1 s$. The upper and lower bound values of the state and input values are $\dot{x}_t \in [3, 10]$ and $\ddot{x}_t \in [-3, 2]$. 

\begin{figure}
    \centering
    \includegraphics[width=0.43\textwidth]{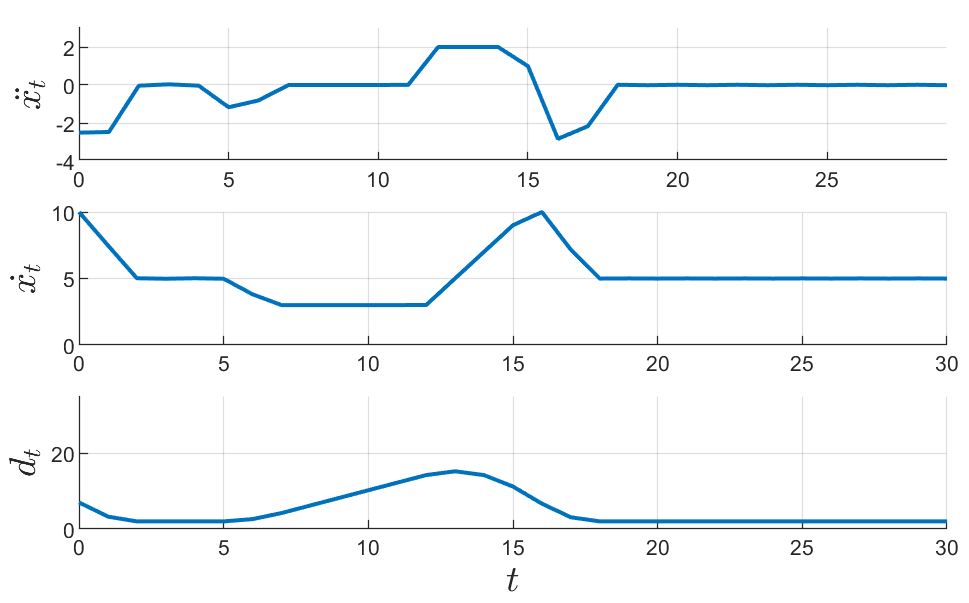}
    \caption{Optimal control input $\ddot{x}_t$, state $\dot{x}_t$ and inter-vehicular distance $d_t$ over the time horizon.}
    \label{fig:stat}
\end{figure}

The trajectory of the lead vehicle is depicted by a red line in Figure (\ref{fig:traj}), and we would like to design a controller that finds an optimal trajectory for the ego vehicle that not only minimizes the conventional motion cost, but also minimizes the expected delay time in the packet delivery (recalling again that the controller knows nothing about the contours in Figure ~(\ref{fig:traj}) in advance). The blue line shows the optimal trajectory generated by GP-MPC for the ego vehicle. It can be seen from the second Figure (\ref{fig:stat}) that, to maintain adequate wireless channel status (and to optimally balance control cost, safety, wireless performance, and do this as fast as possible), the ego vehicle decelerates to increases its distance from the lead vehicle. This policy sacrifices a bit of short-term performance in order to maintain connectivity, improving situational awareness and ultimately long-term performance.

\section{Conclusions}\label{sec:conclusions}
In this paper, an efficient  Model Predictive Control algorithm based on Gaussian Processes is presented to account for uncertainty in the communication channel of motion policies in connected vehicle applications. The presented algorithm learns the wireless channel model in terms of expected packet delivery time by leveraging Gaussian Processes. Due to the substantial amount of available training data, the obtained GP model is very large and computationally expensive to deal with. 

To solve this problem, the algorithm focuses on the \textit{N-step reachable set} from the current state of the ego vehicle as the useful training data set. This decreases the size of the required training data for the current state of the vehicle dramatically. Subsequently the resultant kernel matrix would be substantially smaller than the original Gram matrix, which needs less computational effort to be inverted and multiplied. In addition, because a major part of the computation involved in GP is conducted to find the inverse of the kernel matrix, the controller exploits the recently calculated and readily available inverse of the kernel matrix from the previous state. We use Sherman-Morrison formula and Schur complement to find the inverse of the current kernel matrix after updating the training data set. These steps decrease the running time to find the expected packet delivery time of the wireless channel for the given time horizon $N$ from $O(n^3)$ to $O(\nu^3N^2)$ where $n$ and $\nu$ are the number of overall training data and drawn data for a single time step, respectively, and $\nu N\ll n$. To demonstrate the approach, a simulation of a leader-follower scenario for two connected autonomous vehicles is developed and the results of the algorithm presented. 

\bibliographystyle{IEEEtran}
\bibliography{main.bib}

\end{document}